\documentclass[%
 aip,
 jap,
 floatfix,
 amsmath, amssymb,
 reprint,%
]{revtex4-1}

\usepackage[utf8]{inputenc}

\usepackage{graphicx}
\graphicspath{ {./figures/} }
\usepackage{dcolumn}
\usepackage{bm}

\begin{document}

\preprint{AIP/123-QED}

\title[Grain growth in Pt microheaters]{Grain growth in Pt microheaters subjected to high current density under constant power \\}

\author{Ottó Elíasson}
 \affiliation{ 
Science Institute, University of Iceland, Dunhagi 3, Reykjavík IS-107, Iceland 
}%
\author{Gabriel Vasile}%
 \affiliation{ 
Science Institute, University of Iceland, Dunhagi 3, Reykjavík IS-107, Iceland 
}%
\affiliation{ 
National Institute of Research-Development for Cryogenics and Isotopic Technologies, Uzinei 4, Ramnicu Valcea RO-1000, Romania 
}%
\author{Snorri Ingvarsson}%
 \email{sthi@hi.is.}
\affiliation{ 
Science Institute, University of Iceland, Dunhagi 3, Reykjavík IS-107, Iceland 
}%


\date{\today}

\begin{abstract}
When $50$ nm thick Pt microheaters of lateral dimensions $1\times10$~$\mu$m$^2$ are subjected to high
electric power their resistance $R$ rises, as expected. Following an
initial rise however there is a gradual decrement in $R$ while constant electric power
dissipation is maintained. We find that this
lowering in $R$ is accompanied by grain growth in the polycrystalline thin Pt
film of our heaters. This is confirmed by XRD measurements and SEM imaging.
Similar growth in grain size is observed in thin Pt films that are oven-annealed
at high temperatures. Thus we argue that maintaining high power dissipation in a
microheater has the same effect on its material structure as post-annealing. We
observe the in-plane grain size of a 50~nm thick as-grown Pt film/heater to be
$D_\parallel=15$~nm. When post-annealed at a temperature of $T=600^\circ$C
for 30 min, $D_\parallel=30$~nm, compared with when electric current is run through a
heater we estimate the mean crystalline length to be $D_\parallel=35$~nm.
\end{abstract}

\pacs{Valid PACS appear here}
\maketitle

\section{\label{sec:inng}Introduction}
Nano- and micropatterned metal wire structures have attracted significant attention due to
their possible applications, which, despite their simplicity might be as
varied as serving as microscopic light sources\cite{Hamann:2008,OttonelloBriano:2014gf,OttonelloBriano:2016gi}, being used in near-field
microscopy \cite{Cvitkovic:2006}, as IR detectors or microbolometers\cite{Renoux:2011kd}
or as sources with tunable spectra.\cite{Chan:2006} Reliability and durability of such
metallic wire structures are very important issues in device applications.  For example in
the case of nano- and microsized structures
\cite{Black:1969fc,Gardner:1987ky,Liew:1989ds,Clement:2001bu} they can fail due to Joule
heating and electromigration\cite{Lloyd:1997tk,Lloyd:1999tz,Banerjee:2001cn,Ceric:2011iw}.
There have been several reports on electrically excited (heated) Pt microstructures and
their thermal radiation
properties\cite{Ingvarsson:2007vga,Au:2008hy,Renoux:2011kd,Vasile:2012}. In the paper we discuss the effects of electrical current annealing of thin film Pt wires. The application of low current density annealing via Joule heating is a known technique \cite{knobel_Joule_1995}. In particular it has been used to affect magnetic anisotropy and domain structure in amorphous ferromagnetic nanowires\cite{sossmeier_Comparison_2007} with current densities in the range of 10$^3$~A/cm$^2$. High current annealing is not as common, perhaps due to risk of electromigration, but has for instance recently been applied to reduce the resistance of an Ag nanowire network \cite{kholid_Multiple_2016}. In their work most of the resistance reduction is caused by welding of junctions between nanowires, improving their electrical contact. Here we go close to the breaking point of the wire (due to electromigration), so we work with really high current densities on the order of 10$^7$~A/cm$^2$. A novelty in our measurements is also the power regulation, which is essential to maintain a constant temperature of the wire.

This article focuses on microscopic Pt wires, that we shall henceforth refer to as \emph{heaters} or
\emph{microheaters}, of in-plane dimensions $1\times 10$~$\mu$m$^2$ (width and
length respectively).  We have found that
the thermal radiation spectrum from heaters of these dimensions is unaffected by their
geometry (in our measurement range), i.e.\ is a blackbody spectrum (depending solely on
the body's temperature).  Thus it is very important to control and regulate the individual
heater's temperature. The temperature of the microheater, $T$, is
linearly dependent on the power dissipation in the heater, $P$, via the relation
\begin{align}
	T = T_0 + \frac{dT}{dP} P\quad ,\label{eq:temp_power}
\end{align}
where $T_0$ is the ambient (room) temperature (i.e. the value of $T$ at $P = 0$~mW), and
$dT/dP$ is the thermal impedance of the microheater\cite{Jonsson:2009wb}.  Thus by
regulating the power dissipation in the heater one simultaneously regulates its
temperature. The electrical resistance of the heaters $R$ varies with the temperature
since
\begin{align}
	R = R_0(1 + \alpha \Delta T)\quad , \label{eq:res_temp}
\end{align}
where $\alpha$ is the temperature coefficient of resistance, $\Delta T = T - T_0$
and $R_0$ is the resistance at room temperature. We observe an increase in $R$
with higher $T$, just as expected.

When our heaters are subjected to high current density, which is done to achieve higher
radiation temperatures and shorter wavelengths, they tend to fail due to electromigration.
The time-to-failure $\tau$, of a given interconnection, or heater in our case, is given by
Black's equation\cite{Black:1969fc} and $\tau\sim1/J^2$. We use a value $n = 2$ for the
current density exponent in accordance with ref.~ \onlinecite{Shatzkes:1986}. In our
work we deal with current densities up to $J\simeq7\times10^7$~A/cm$^2$. 

In our research on electromigration and the time-to-failure of Pt microheaters, we have observed a gradual decrease with time in $R$, when microheaters are excited by electric current at constant power\cite{Eliasson:2014}. We know from previous work that such changes in $R$ do not affect the thermal impedance $dT/dP$ as this is governed by the thermal properties of the surroundings, most importantly the substrate's thermal conductivity\cite{Jonsson:2009wb}. This slow drift in $R$ led us to further investigate the metal structure of our heaters by other means than just by measuring their resistance profiles. We hypothesize that subjecting the heaters to high power, and thus high temperature, has the same effect on the material structure as if they were annealed in an oven. 
The observed resistance drop seems similar to the effect of high current annealing reported in \cite{kholid_Multiple_2016}, but in our case the cause is different. Here we attribute the resistance change to a reduction in electron scattering at grain boundaries.

\begin{figure}[t]
	\centering
	\includegraphics[width=0.49\columnwidth]{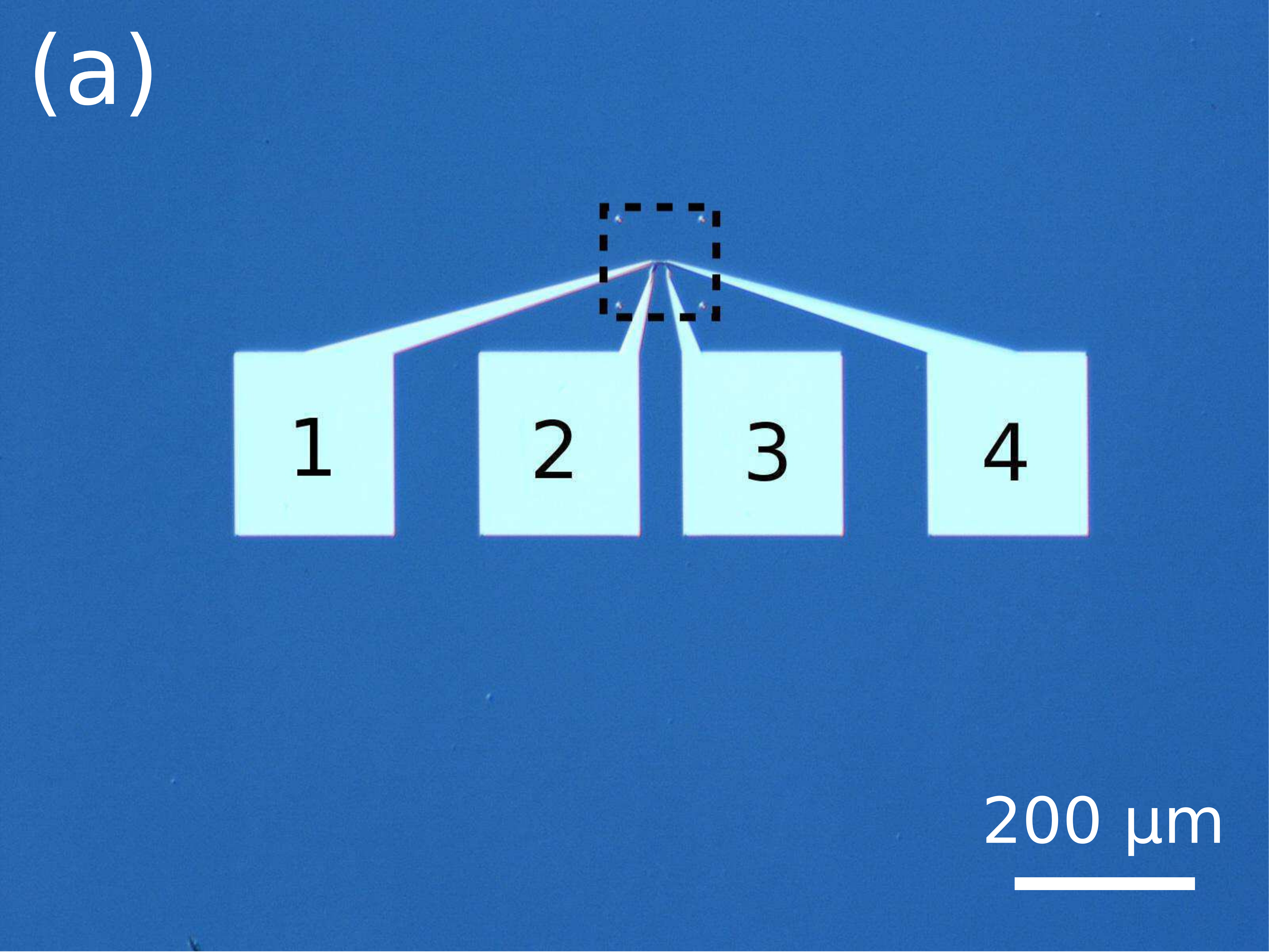}
	\includegraphics[width=0.49\columnwidth]{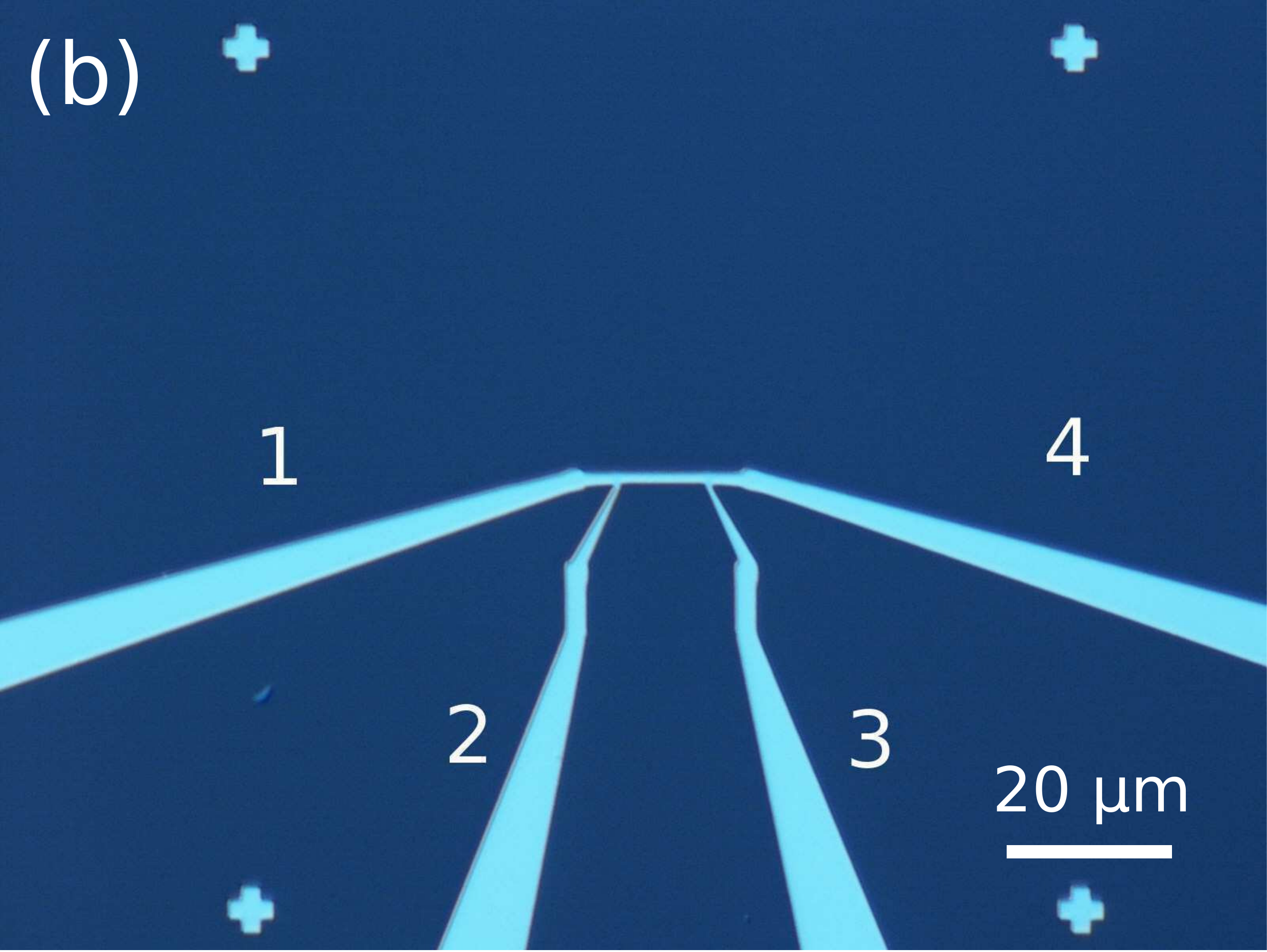}
	\caption{Heaters used in the research. The terminals are labelled with numbers
	1--4 and each is about 200$\times$180~$\mu$m$^2$.  The dotted line in fig.~(a)
demarcate the area pictured in fig.~(b). Images were obtained using an optical microscope.}
	\label{fig:heater}
\end{figure}

The outline of this paper will be as follows: At first we introduce our the experimental
setup and samples. Then we focus on the distinct $R$-$t$ (resistance-time) and $R$-$P$
(resistance-power) profiles of our heaters. In the subsequent section we estimate the grain
size of our heaters by two methods, by Scanning Electron Microscopy (SEM) imaging and by
grazing incidence X-ray diffraction (GIXRD) measurements.

\section{Sample properties}
The heaters used in our research are manufactured by electric beam (e-beam) lithography to
ensure that their shape is well defined. A Si/SiO$_2$ (100~nm oxide) substrate is coated with a
negative e-beam resist. The heaters are patterned in the resist and the part exposed to the beam
is developed away. The heaters, typically about 60 on each substrate chip, are deposited via
sputtering where a 5~nm thick polycrystalline Cr adhesion layer is grown before the 50~nm
polycrystalline Pt layer. The growth rate of Cr is 3.2~nm/min and that of the Pt is 5.2~nm/min. 
For further details on the film growth we refer to \cite{Au:2008hy}.
A lift-off step in an acetone bath leaves us with the final structure.  The chip is rinsed and
then the heaters are ready for use. The dimensions of the microheaters used in this study are
1$\times$10~$\mu$m$^2$, and their thickness is as detailed above. 
We have measured the surface roughness of our films by atomic force microscopy, and in all cases the rms roughness is below 1~nm.
A typical microheater is depicted in fig.~\ref{fig:heater}.

We use four terminal electrical measurements to obtain the power dissipation in our heaters,
$P$, as this enables us to have accurate control over the temperature of the heaters.  We use a
homemade circuit to both monitor and to regulate the power dissipation in the heater. This
process is computer controlled via a programmable National Instruments Data Acquisition board
(DAQ-board, NI-USB 6229).  The setup and the circuit is thoroughly described in
\cite{Eliasson:2014}. We subject our microheaters to a high current density, and wait until
they eventually fail due to electromigration, in order to test their time-to-failure.  By
switching the current polarity at a frequency of $f = 20$~kHz (referred to as AC current stress
in the article) we observed on the order of 10$^3$-fold increase in the time-to-failure of our
heaters, compared to when they are biased at the same power level at DC\cite{Eliasson:2014}.
Depending on the bias, our measurements can take quite some time, the longest measurement on a
single heater lasted for over a week.

Three quantities describing the properties of our heaters are of special interest. These are
their electrical resistance $R$ (in particular the cold resistance $R_c$), the thermal
impedance $dT/dP$ and the temperature coefficient of resistance $\alpha$. Attempts were made to
determine if any permanent changes occur in these quantities upon subjecting our heaters to
high power.  Our findings show that both $\alpha$ and $dT/dP$ appear to be essentially
unchanged despite thermal cycling, whether it be by varying the current density or by external
heating (on a hotplate). However, we found that the heater resistance changes substantially in
our samples and devote section~\ref{sec:rtgraphs} to discussion on those observations.

These properties, $\alpha$ and $dT/dP$, can be obtained by current-voltage (\mbox{$I$-$V$})
measurements, with the help of a hot plate to vary temperature, as follows:  An as-grown heater
is \mbox{$I$-$V$} characterized. These measurements result in a cold resistivity of  
\begin{align}
	\rho_c = 20 \ \mu\Omega\text{ cm}. \label{eq:resist}
\end{align}
This value is about 90\% greater than the bulk resistivity of Pt\cite{kittel}, $\rho_0$~=~10.4 $\mu\Omega$~cm, but it should be kept in mind that this is a thin and narrow
wire with large contribution to resistivity from surface scattering. Its $R$-$P$ profile
is stored likewise. In the linear part of the $R$-$P$ graph the line is described by combining
eqs. (\ref{eq:temp_power}) and (\ref{eq:res_temp}) \begin{align} R = R_0\left(1 + \alpha
	\frac{dT}{dP} P\right)\label{eq:RP} \end{align}
where $P$ is the power dissipated in the heater and $R_0$ is the resistance at $P = 0$~mW. 
\begin{figure}[ht]
	\centering
	\includegraphics[width=0.49\textwidth]{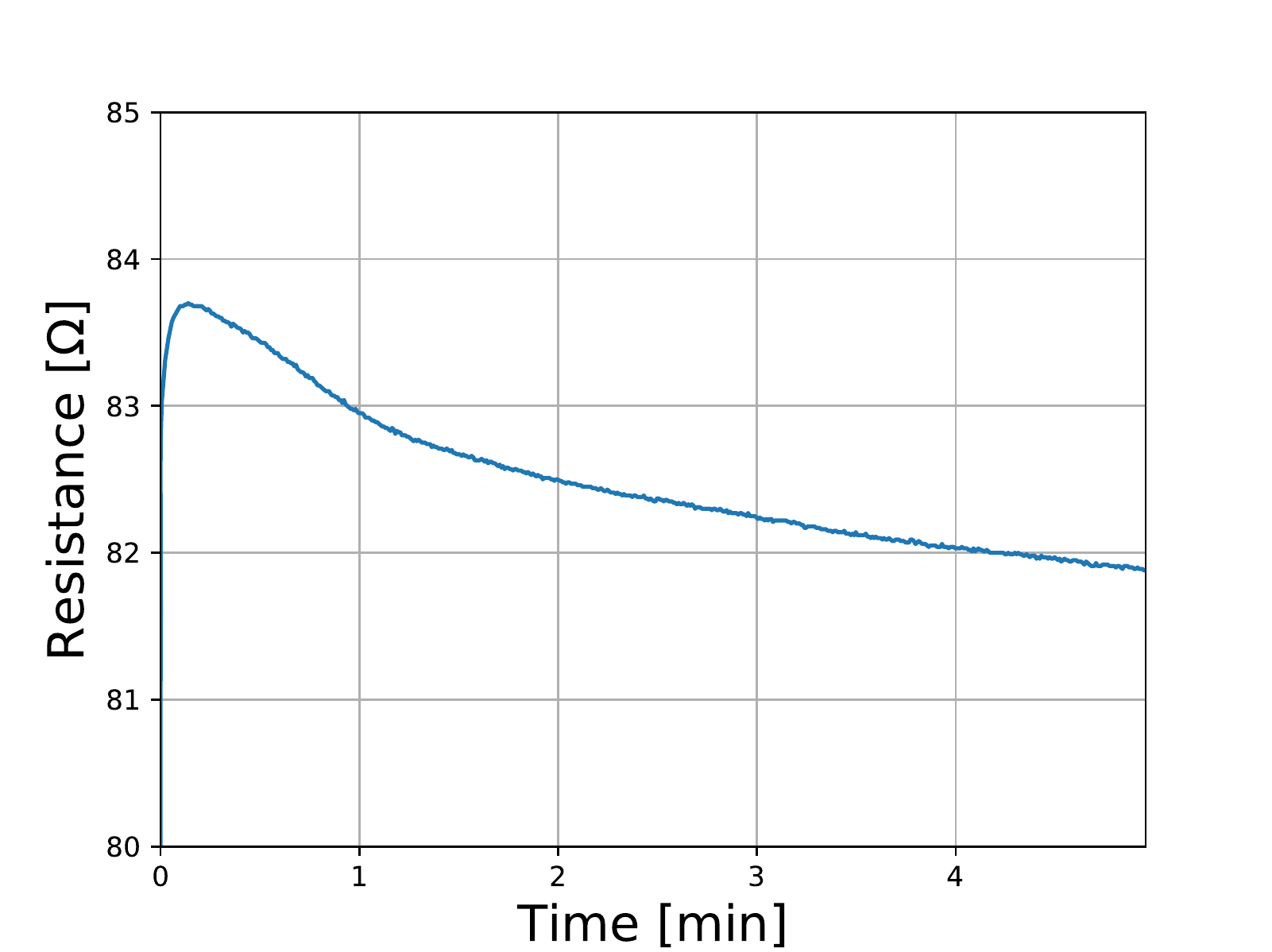}
	\caption{$R$-$t$ graph of a microheater used in the research set on $P_{\rm h} = 90$ mW with AC current stress.}
	\label{fig:res_time}
\end{figure}
The heater is now measured on a hot plate, whose temperature is increased stepwise and changes in resistance are observed. We measured over a temperature range from room temperature up to about 110$^\circ$C, and found $\alpha$ from fitting eq. \ref{eq:res_temp} to be
\begin{align}
	\alpha = 2.20\times10^{-3} \text{ K}^{-1}.
\end{align}
From the $R$-$P$ graph the value of $dT/dP$ can now be estimated. For our heaters we obtained a value of
\begin{align}
	\frac{dT}{dP} = 4.7 \text{ K/mW}. \label{eq:dTdP}
\end{align}
The heater is now subjected to constant high power (which is the quantity regulated) for a few minutes, a time long enough so an obvious decrease in the heaters resistance has been observed, see sec.~\ref{sec:rtgraphs}.  Following this procedure the heater's $I$-$V$ profile is obtained again to measure the cold resistance and its $\alpha$ value is measured again to see if it had changed. We found no changes in $\alpha$ or $dT/dP$ as mentioned earlier.
This is in agreement with results from Pt films annealed to temperatures of 1300$^\circ$C\cite{zhang_Microstructure_1997}.

All the heaters in our study are manufactured in the exact same manner, therefore we assume
they have the same physical characteristics.  We chose to regulate heater power at a value of
$P = 90$~mW in our measurements.  This power value was obtained by trial and error, as heater
failure is a statistical event.  It was chosen such as to see a substantial decrease in $R$ in
the first few minutes upon biasing and a ``reasonable'' time-to-failure. At this high power we
estimate the temperature in light of eq.~(\ref{eq:temp_power}) and (\ref{eq:dTdP}) as being
$T_{\rm h} = 710$~K corresponding to about 440$^\circ$C.

\section{Monitoring resistance}\label{sec:rtgraphs}
The effect of temperature on the resistance of a metal slab is well known and may be described
by eq.~(\ref{eq:RP}). When subjected to high power, the resistance of our heaters rose during
the first few seconds as can be expected due to self-heating, but when the desired regulation
power (90~mW) was reached we noted a decrease in $R$---steep at first but then with a gradually
declining slope, as if the resistance follows an inverse power law. This can be seen in
fig.~\ref{fig:res_time}, that shows an $R$-$t$ (resistance-time) graph of an as-grown heater
whose power was regulated at $P=90$~mW under AC current stress. 
In this particular measurement we ramped the heater power up slowly and then ramped it back down after 
5~minutes of constant high power regulation. In ref. \onlinecite{Eliasson:2014} there can be found 
examples of measurements where after the drop, the resistance reaches a minimum, followed 
by a rise in the resistance leading to the eventual break of the heater due to electromigration.

This early drop in $R$ is also observed in as-grown heaters subjected to DC current stress. If
the bias is maintained for long enough time the resistance typically starts to rise again and
eventually the heater breaks due to electromigration\cite{Eliasson:2014}.  Further, it appears
to reflect an irreversible change in the sample properties, i.e.\ the material structure, as far as we can see. 

\begin{figure*}[t] 
	\includegraphics[width=0.49\textwidth]{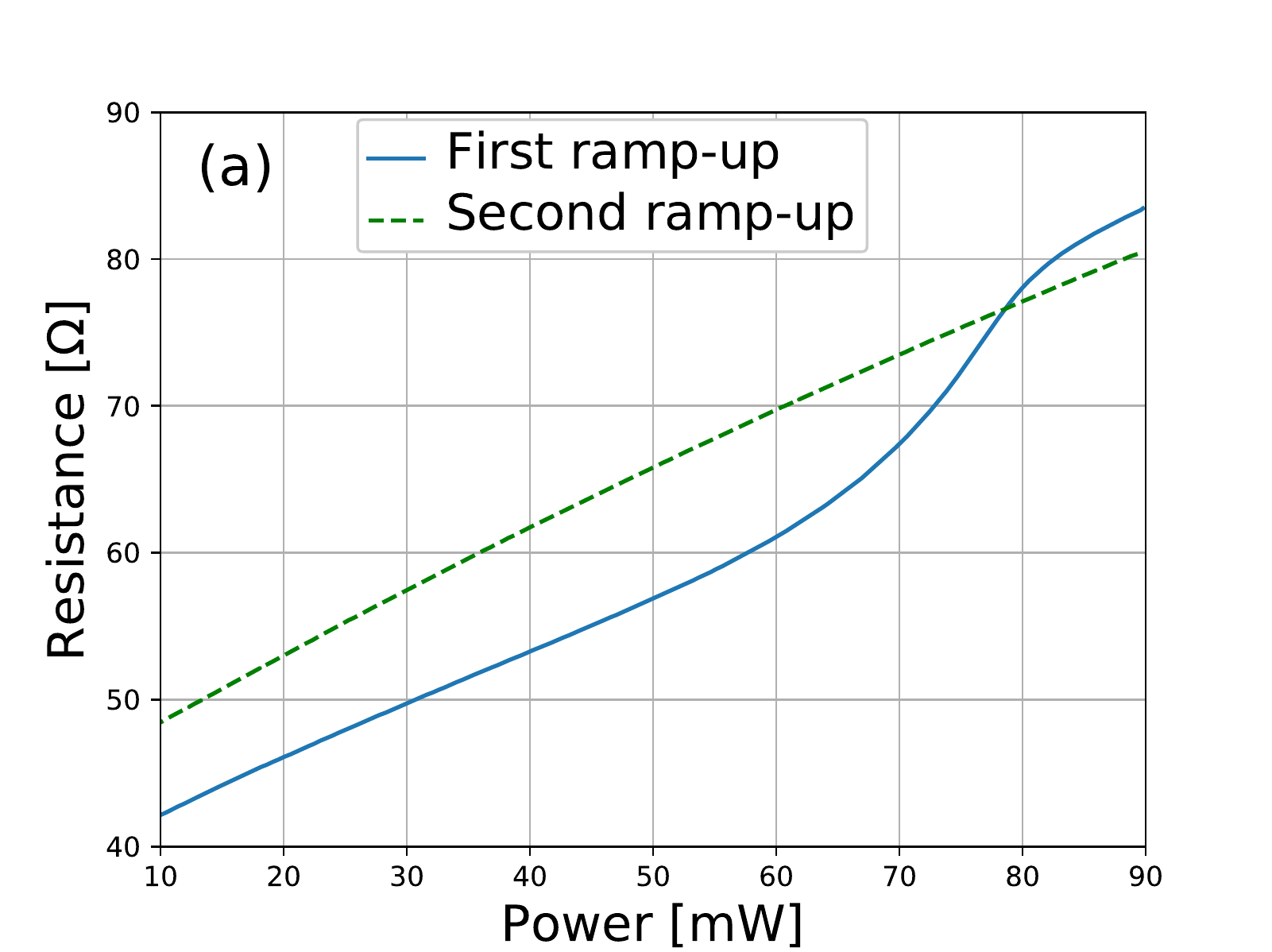} 
	\includegraphics[width=0.49\textwidth]{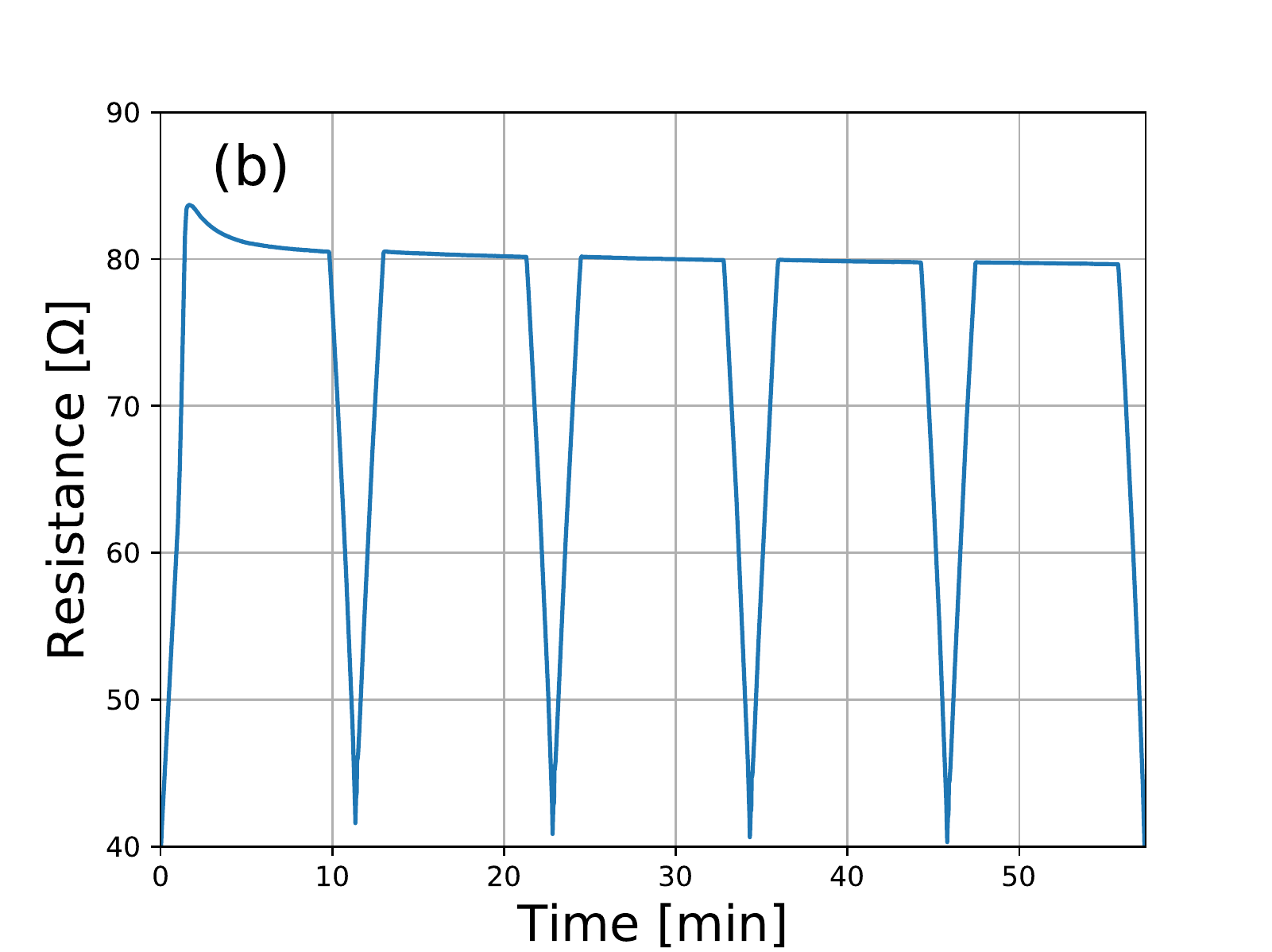}	
\caption{Ramp-up/-down measurements of a microheater.  Power was ramped up to 90~mW and down to 1~mW, five times at regular intervals.  (a) Resistance-power ($R$-$P$) graph of a microheater during the first and second ramp-up of power.  (Subsequent ramp-up curves are almost identical to the second one and are omitted for clarity). Notice the steep rise in resistance around 70~mW (corresponding to about 370$^\circ$C) on the first ramp-up curve.  (b) Resistance-time ($R$-$t$) graph of the same process. Most of the resistance drop takes place during the first cycle in this case.}\label{fig:ramp} 	
\end{figure*}

An interesting effect can be seen in fig.~\ref{fig:ramp}~(a), which is an $R$-$P$ graph of a
heater where the power is ramped up at 1~mW/s. It compares the resistance during the first two
ramp-ups, which are quite distinct. Later ramp-ups (the 3rd, 4th etc.) were almost identical
to the second one, and thus omitted in the graph.  During the first ramp-up there is a steep
rise in the resistance at $P\simeq70$~mW (power is maintained constant in 1~mW steps). This
power value corresponds to a temperature of about 370$^\circ$C, according to eq.
(\ref{eq:temp_power}). We take this as a sign of permanent change in the heater's material
structure which seems to happen when its power is driven above a certain limit for the first
time. A reduction in $R$ is caused by less electron scattering, which we attribute to growth in
grain size of the polycrystalline Pt film\cite{Zhang:2006ct,Wang:2011jg}. 

Fig.~\ref{fig:ramp}~(b) depicts an $R$-$t$ graph of the same heater, again with the power
ramped up at a rate of 1 mW/s. When the desired power is reached it is maintained for several
minutes, and then lowered again (corresponding to the dips in resistance). This is repeated at
regular intervals. As can be seen in the figure, the resistance trace reaches the same value after each ramp-up as it had prior to the ramp down of the power, further supporting the hypothesis of irreversible
change to the sample.  

Changes in $R$, imply changes in the resistivity $\rho$
of the heater. Based on
Matthiessen's rule,
we can simply add the contributions to the resistivity of a thin film $\rho_{\rm f}$, such that
\begin{align}
	\rho_{\rm f} = \rho_{\rm 0} + \rho_{\rm GB} + \rho_{\rm SS} + \rho_{\rm SR}.
	\label{eq:res_model} \end{align}
Here $\rho_{\rm 0}$ is the bulk resistivity, $\rho_{\rm GB}$ a term caused by grain boundary
scattering, $\rho_{\rm SS}$ comes from surface scattering and $\rho_{\rm SR}$ results from the
surface roughness of the film. \cite{Wissmann:2007tv} Our films are smooth
compared to the thickness, so we can safely say that $\rho_{\rm SR}$ is very small and will not be
considered here. According to the Fuchs-Sondheimer model\cite{Fuchs:1938tn,Sondheimer:1952fz}
$\rho_{\rm SS}$ is a constant quantity for a constant value of the thickness $d$ of the metal
film. 
We argue, based on measurements presented in the following section, that the main reason for the reduction in $\rho_{\rm f}$ observed in fig. \ref{fig:res_time}, is due to grain growth in the Pt film, which yields a decrement in $\rho_{\rm GB}$.

\section{Grain size measurements} \label{sec:grainsize}
In order to test the grain growth hypothesis we conducted two kinds of measurements to measure
the grain size of the heaters and of Pt thin films grown in an identical manner to the ones
our heaters are patterned from. We employ grazing incidence X-ray diffraction
(GIXRD) measurements and scanning electron microscope (SEM) which give different
information about the sample. 
GIXRD measurements are useful for polycrystalline thin films where the grains are randomly oriented. One also typically obtains better signal strength than in ordinary $\theta$-2$\theta$ measurements. It should be kept in mind that during GIXRD the incident X-rays are kept at a fixed low angle while the detector is scanned through a range of angles. As a result, the different peaks in the GIXRD scan represent different plane orientations (or $k$-vector directions).
While SEM imaging yields an estimate of $D_\parallel$, different GIXRD peaks can be used to obtain estimates of grain sizes in directions corresponding to the $k$-vector orientation.
Further, we
compare the effect of running large electric current (high current density) in our samples with
that of annealing identical thin films in an oven at various temperatures.  As mentioned
earlier the self heating at high current density raises the heater temperature quite
significantly (440$^\circ$C at 90~mW power dissipation).

\begin{figure*}[t] 
	\includegraphics[width=0.49\textwidth]{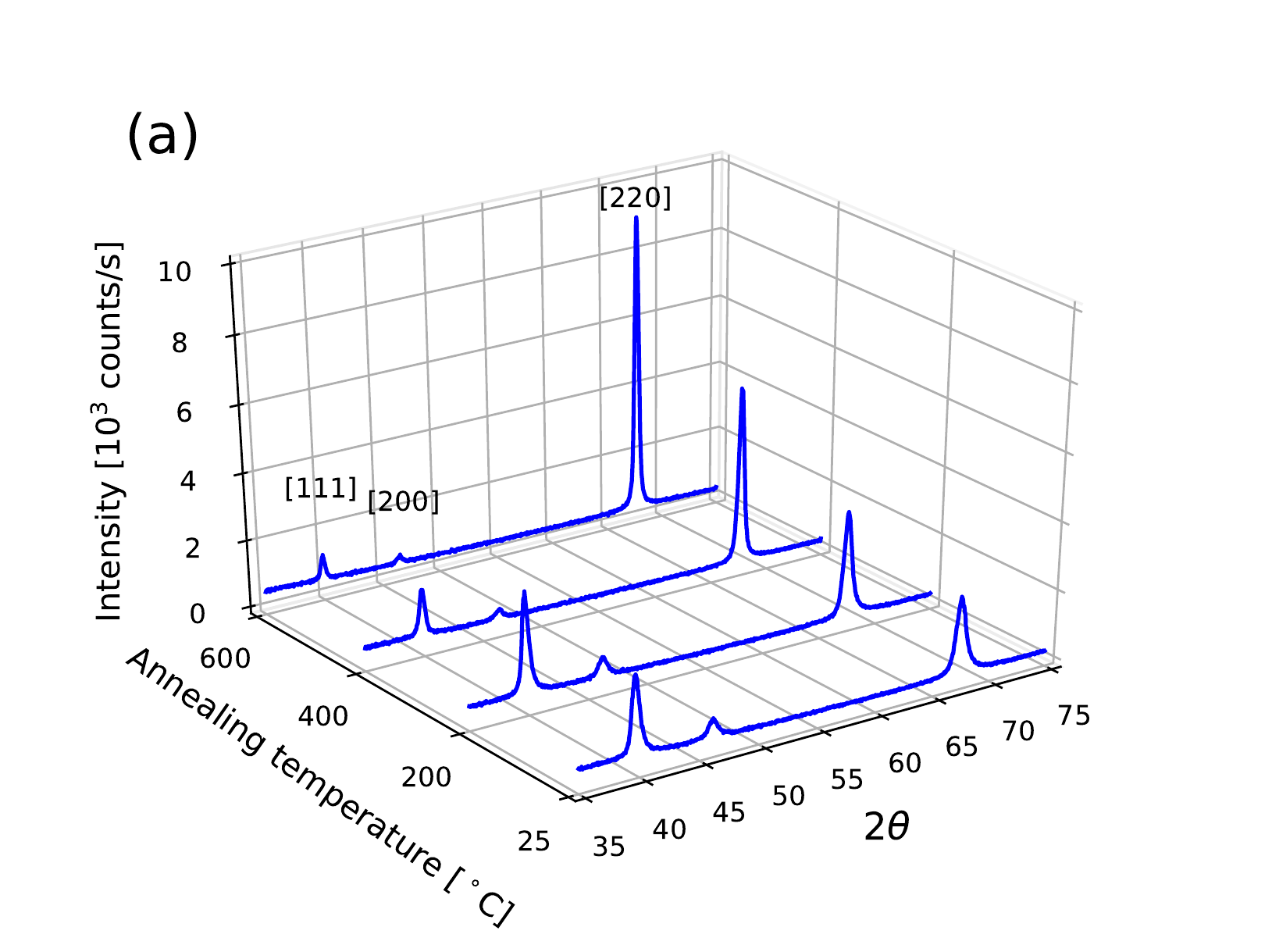}
	\includegraphics[width=0.49\textwidth]{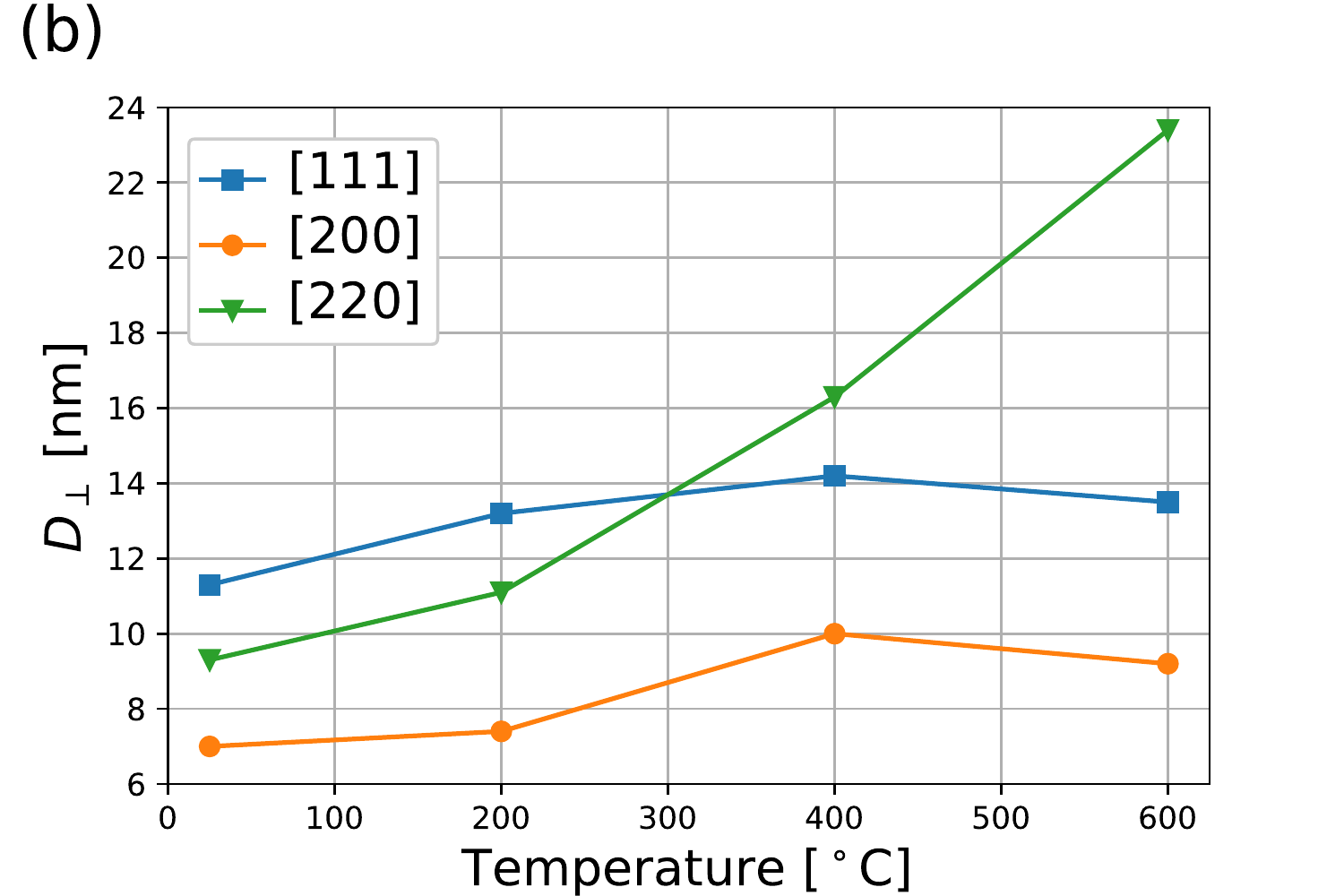}
\caption{(a) Grazing incidence X-ray spectra of Pt films. The films were annealed for 30 minutes, apart from the one closest to the bottom of the graph, which is unnannealed. Peaks corresponding to crystalline planes [111], [200] and [220] are labelled.
  (b) The grain size estimates from GIXRD peaks increase with increasing post annealing temperature.} \label{fig:temp_grain_GIXRD}
\end{figure*}

Our heater structures are too small to measure them directly in an X-ray apparatus. Thus four 
Cr/Pt films were grown in the same manner as the heaters, and three of these were post-annealed
for 30 minutes at 200$^\circ$C, 400$^\circ$C and 600$^\circ$C temperature, respectively.  The
results of subsequent GIXRD measurements on all four films are presented in fig.~\ref{fig:temp_grain_GIXRD}~(a).
Using the Scherrer equation, $D_{GI}$ (crystalline coherence length), can be calculated in the
following manner:\cite{Langford:1978te}
\begin{align}
	D_{GI} = \frac{K\lambda}{\beta\cos(\theta)}.
\end{align}
Here $\beta$ is the FWHM of the peaks in the GIXRD spectrum (three peaks were observed
corresponding to crystalline planes [111], [200] and [220]), $\theta$ is the peak position in
the spectrum, $\lambda$ is the wavelength of the X-rays used and $K$ = 0.9 is the Scherrer
correction factor. Scherrer's equation is used to compare values of $D_{GI}$ and observe
trends, rather than for exact estimation of the grain size perpendicular to film plane. We graph the results in fig.~\ref{fig:temp_grain_GIXRD}~(b). Grain growth is apparent with increasing post-annealing temperature.  

Similar measurements on Pt were not found in the literature, but in
ref.~\onlinecite{Wissmann:2007tv}, GIXRD measurements on thin Ag films are discussed. Those
results are in many respects identical to ours. They were conducted at a lower temperature
range from about 90 K, up to 500 K (see chapter 3 of ref.  \onlinecite{Wissmann:2007tv}). In
fig.~\ref{fig:temp_grain_GIXRD}~(b) we see that the [111] and [200] peaks yield slowly increasing
values of $D_{GI}$, but appear not to undergo substantial changes. This is in accordance
with ref.~\onlinecite{Wissmann:2007tv}, where those peaks give a constant value of
$D_{GI}$ for the same temperatures as ours, from 300 K and up. In addition we also measure the
peak corresponding to the [220] crystalline plane. These planes appear to grow substantially in
the temperature range of our measurements, just as the [111] and [200] planes did on lower
temperatures in ref. \onlinecite{Wissmann:2007tv}. 

The [220] crystalline planes appears to dominate in the structure, since the
intensity of the peak increases at the cost of the other two planes measured.
One also sees on fig. \ref{fig:temp_grain_GIXRD}~(a), that the intensity of the [220] diffraction
peak starts to grow between 200$^\circ$C and 400$^\circ$C. We associate this change
with the steep rise in electrical resistance occurring during the first ramp-up curve of fig.~\ref{fig:ramp}~(a), at a heater temperature of about 370$^\circ$C. This is an indication of changes in the material structure and at this temperature of 370$^\circ$C the activation energies for these processes have been reached.

\begin{figure*}[ht]
	\centering
	\includegraphics[width=0.49\textwidth]{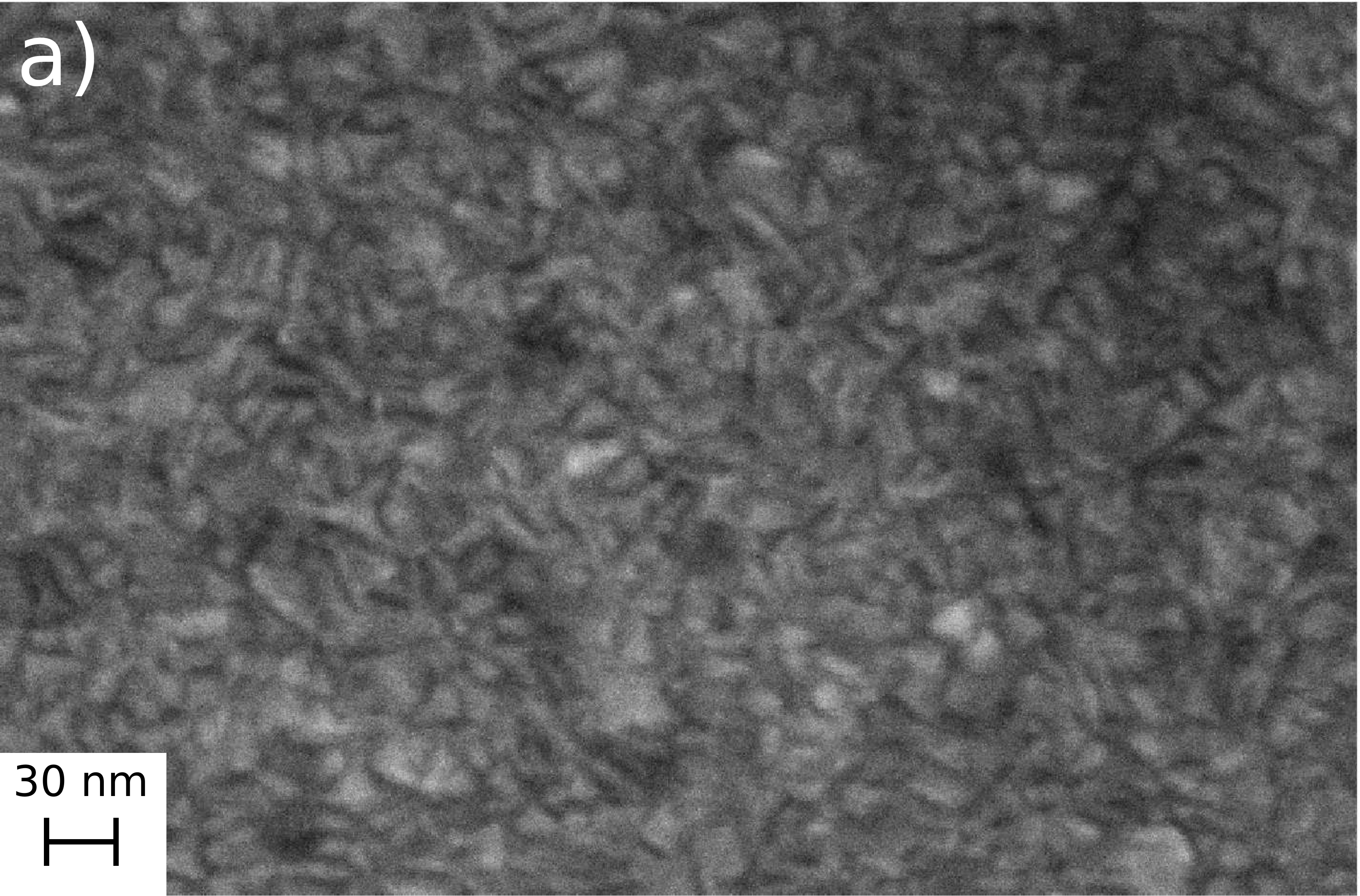}
	\includegraphics[width=0.49\textwidth]{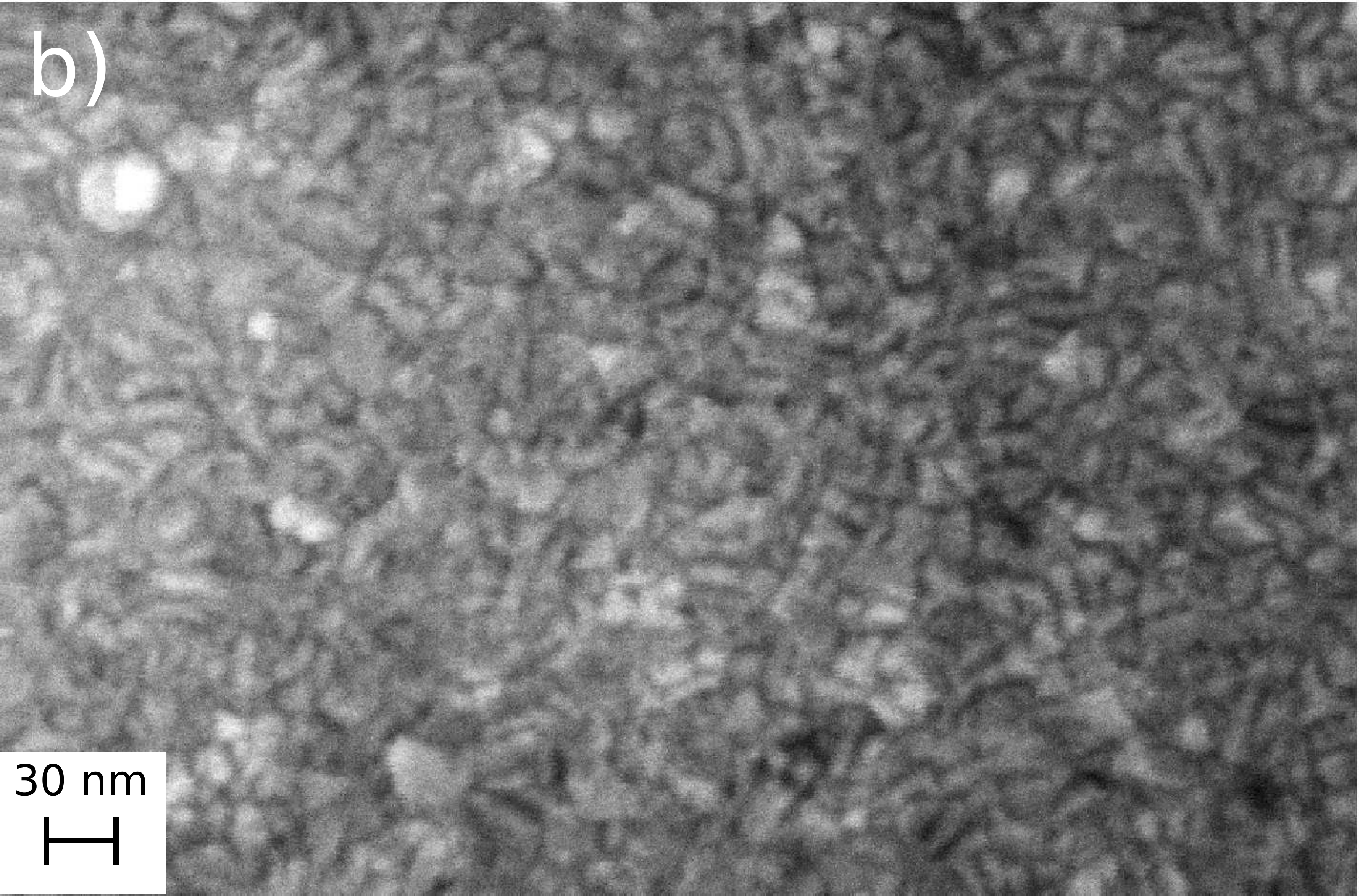}
	\includegraphics[width=0.49\textwidth]{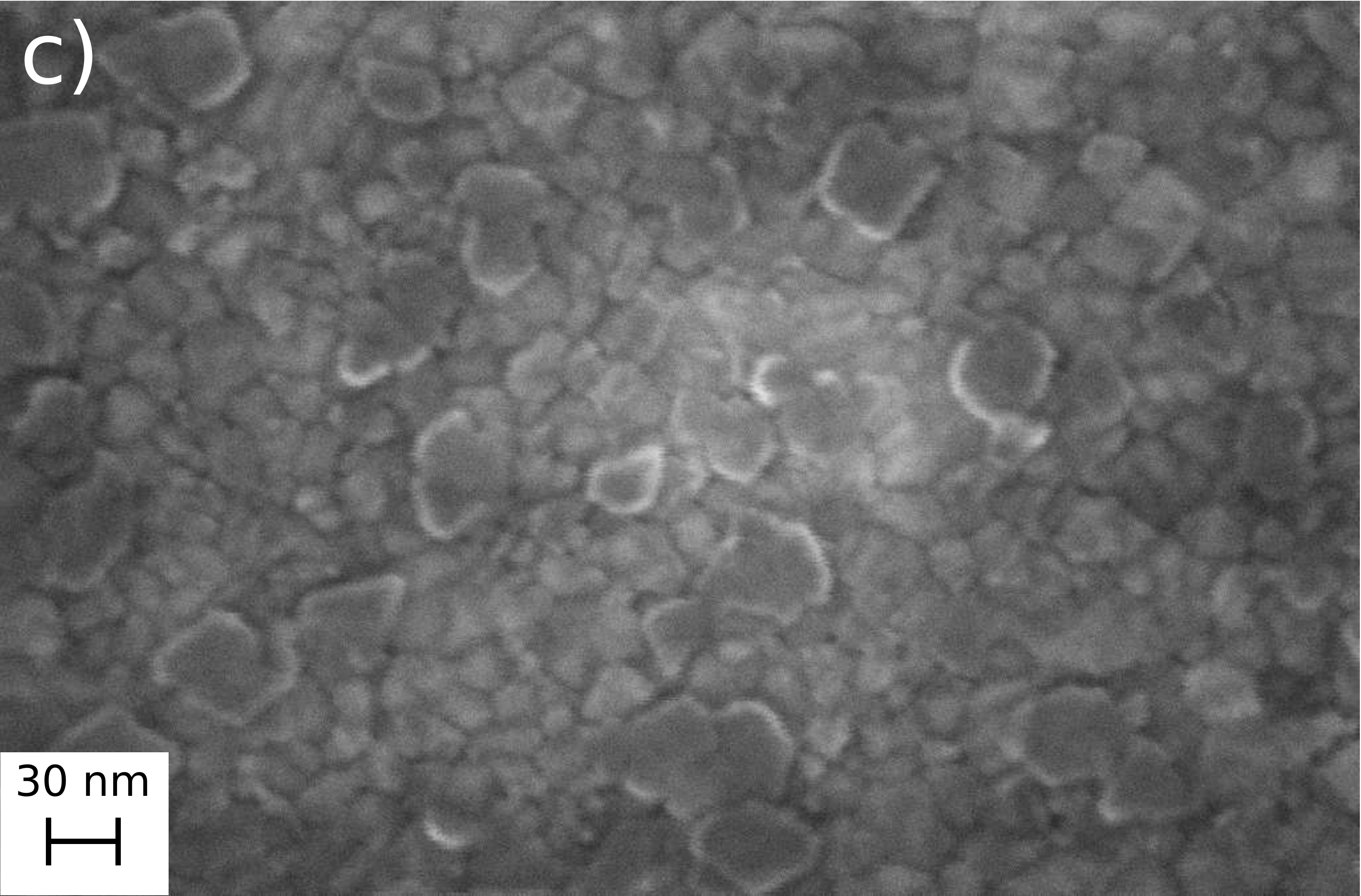}
	\includegraphics[width=0.49\textwidth]{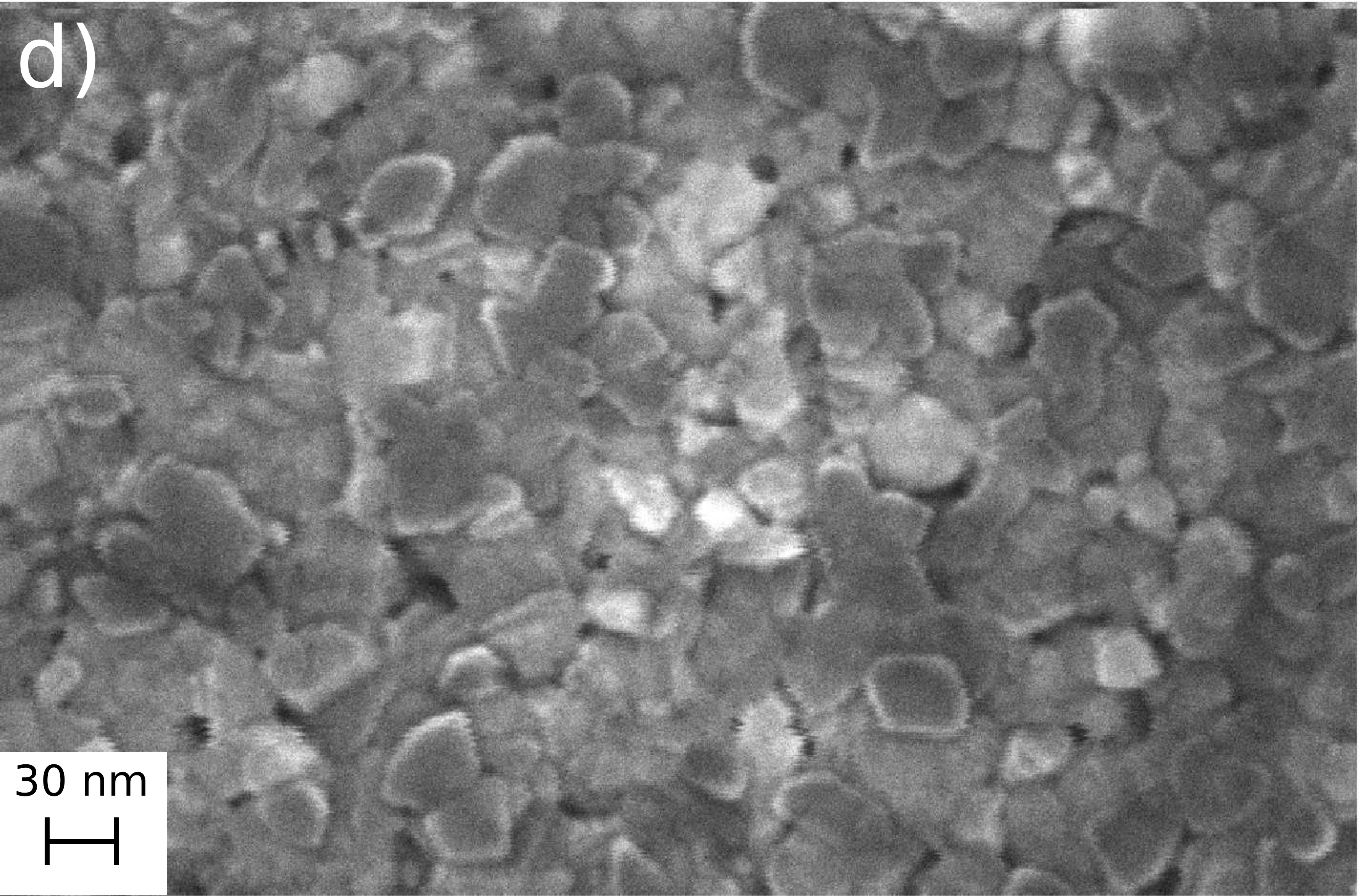}
	\includegraphics[width=0.49\textwidth]{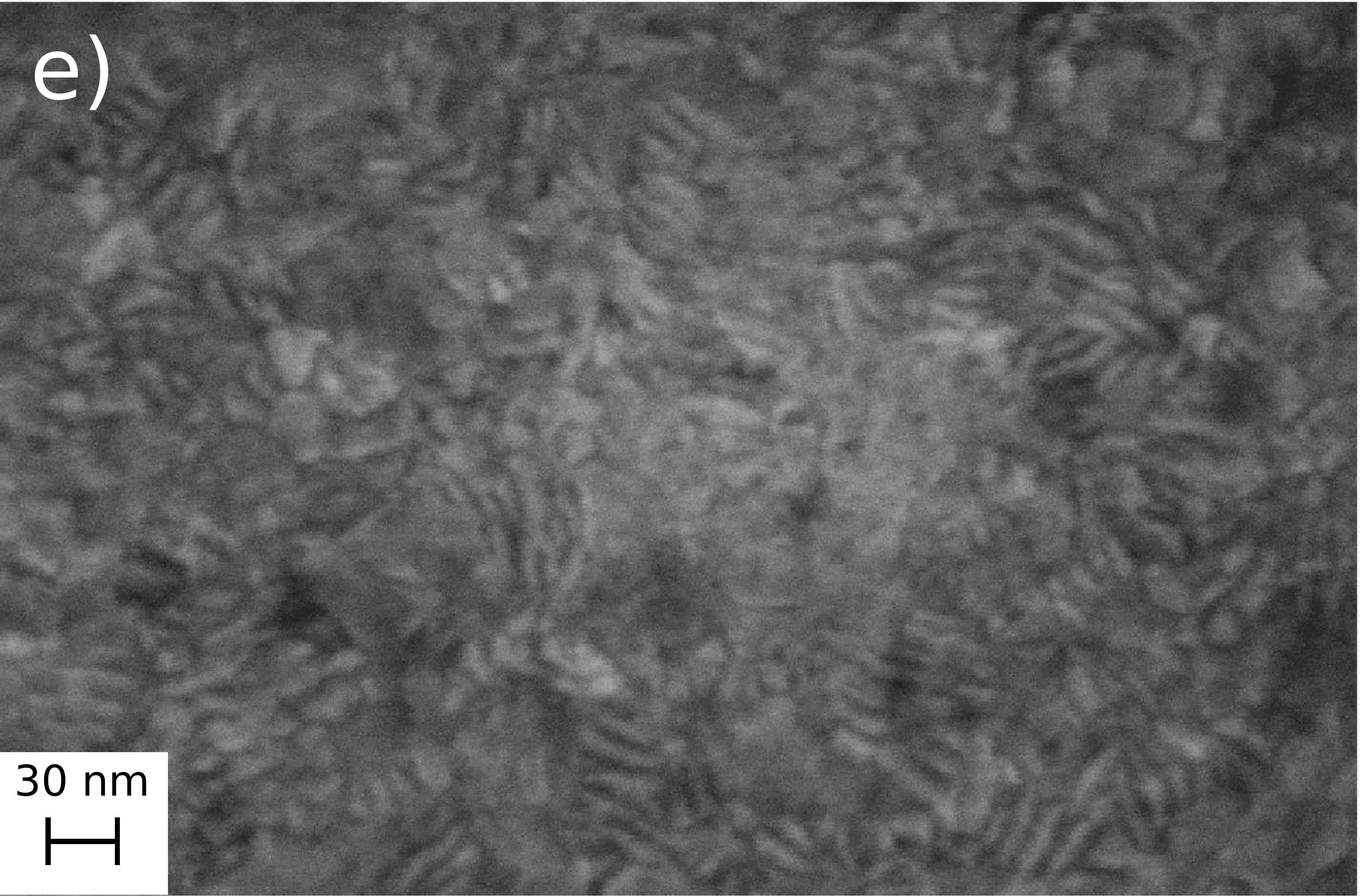}
	\includegraphics[width=0.49\textwidth]{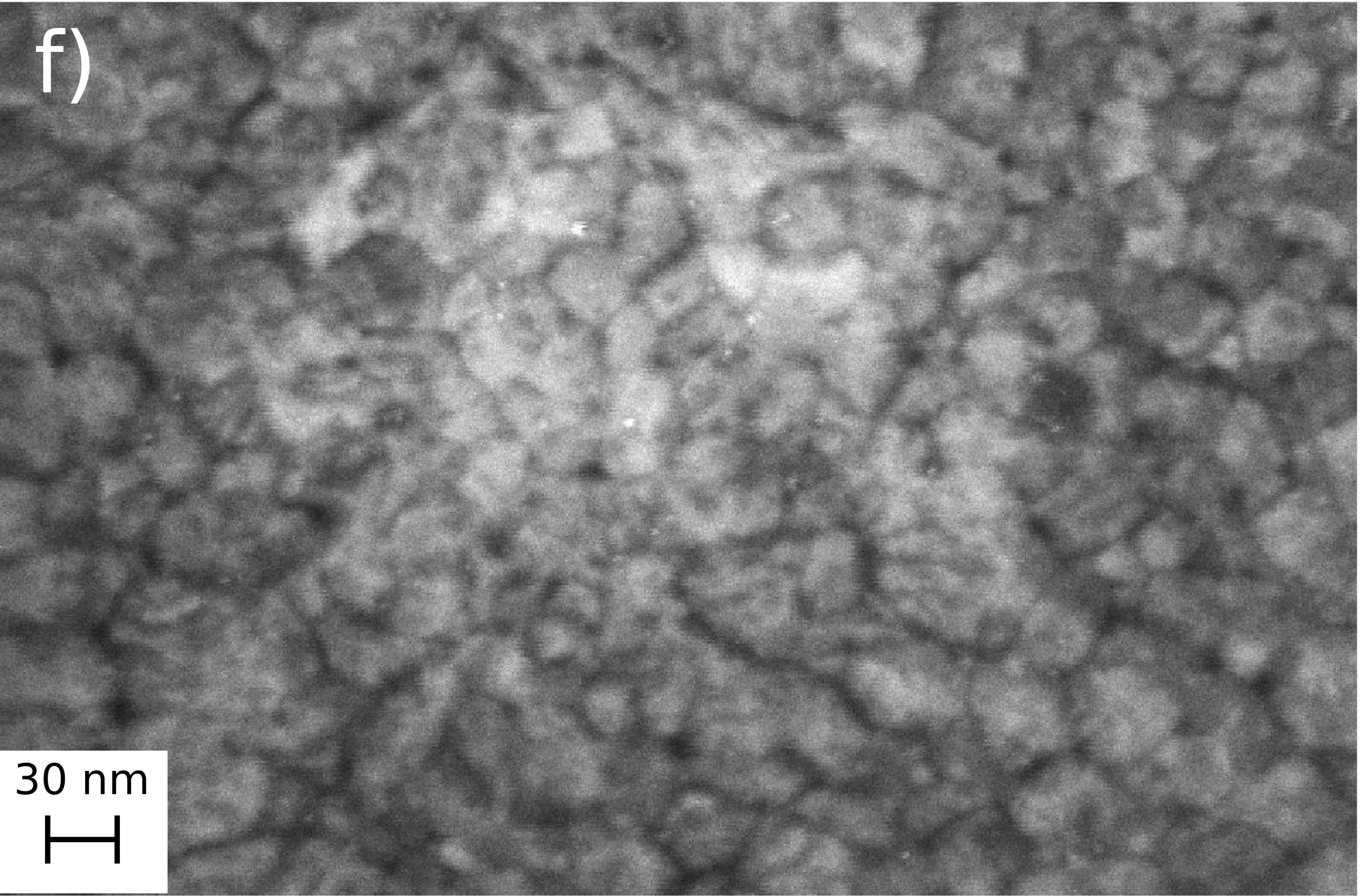}
	\caption{Figures (a)--(d) are SEM pictures of thin films and figures (e) and (f) are SEM pictures of heaters before and after measurement. Grain growth is apparent in both cases.
	(a) Surface of an unannealed 50 nm Pt film. The in-plane grain size is about $D_\parallel\sim$~14~nm.
	(b) Surface of a 50 nm Pt film, annealed at 200$^\circ$C for 30 min. The mean in-plane grain size is $D_\parallel=$~17~nm.
	(c) Surface of a 50 nm Pt film, annealed at 400$^\circ$C for 30 min. The mean in-plane grain size is $D_\parallel=$~26~nm.
	(d) Surface of a 50 nm Pt film, annealed at 600$^\circ$C for 30 min. The mean in-plane grain size is $D_\parallel=$~30~nm.
	(e) Surface of a Pt heater before electrical measurement. Mean in-plane grain size is  $D_\parallel=$~16~nm.
	(f) Surface of the same Pt heater as in (e) after electrical measurement. Mean in-plane grain size is  $D_\parallel\sim$~35~nm.
	}
	\label{fig:pt_ann}
\end{figure*}

The in-plane grain size $D_\parallel$ was also estimated in a SEM, see figs.~\ref{fig:pt_ann}~(a)--\ref{fig:pt_ann}~(f). Grain growth is apparent upon annealing.  $D_\parallel$ varies from about 15~nm for the unannealed film up to 30~nm for the the one annealed at $600^\circ$C. In addition figs.~\ref{fig:pt_ann}~(e) and \ref{fig:pt_ann}~(f) display the grain size of microheaters before and after electrical measurement at 90~mW constant power. The same grain growth happens when the heater is subjected to high current density which raises its temperature according to eq.~\ref{eq:temp_power}. The material is annealed, just not in an oven but by self-heating. In addition one notices that not only the size of the grains changes, but also the shape. The grains appear to take on a rounder shape in contrast to the more oblong shape of the grains in the unannealed films. Similar growth of post-annealed thin Pt films is reported in ref.~\onlinecite{Agustsson:2008dr}.

\section{Conclusions}
We investigated $50$ nm thick Pt microheaters of lateral dimensions  1$\times$10 $\mu$m$^2$, subjected to
current density up to about $J\simeq7\times10^7$~A/cm$^2$. We regulate the electric power dissipated in
the heater at a constant value of
$P_{\rm h} = 90$~mW, obtained by trial and error, chosen to see a substantial
decrease in $R_{\rm h}$. At this high power we estimate the temperature in light of eqs.~(\ref{eq:temp_power}) and (\ref{eq:dTdP}) to be $T_{\rm h} = 710$~K, about
440$^\circ$C. As expected, by increasing the power dissipation in the microheater
the resistance increases as described by eq.~(\ref{eq:RP}). When the desired regulation power
was reached, the heater's resistance drops, steeply at first but then the process slows down. The decrement in the heater resistance is permanent, which indicates
a modification of the heaters' structural properties. To investigate the structure of these heaters, we used
two different techniques: SEM and GIXRD. From those measurements we see that the grains in thin Pt films start to grow when heated to temperatures between 200$^\circ$C and 400$^\circ$C. Indeed we also see signs of these structural changes in the heater when current is run through it the first time, then at heater temperature of about 370$^\circ$C. We conclude that the observed decrease in  the heaters resistance, is due to grain growth in the metal as it is subjected to high electrical power.


\begin{acknowledgments}
This work was partially supported by the Icelandic research fund.
\end{acknowledgments}

\bibliography{eliasson}
\end{document}